\documentclass[10pt,reqno]{amsart}

\usepackage[margin=0.625in,bottom=1in,top=0.75in,twosideshift=0in]{geometry}
\usepackage{multicol}
\newcommand{\mysubsection}[1]{\subsection{\normalfont \text{ }\emph{#1}}\text{}\indent}
\usepackage{amssymb, amsmath}
\usepackage{amsthm}
\usepackage[usenames]{color}
\usepackage{calc}
\usepackage{times}
\newcounter{myFCounter}[section]
\usepackage{graphicx}
\newcommand{\myFigure}[4]{
    \footnotesize
    \begin{center}
    \begin{minipage}[!t]{\columnwidth}%
        \begin{center}\refstepcounter{myFCounter}\vspace{1ex}%
        \includegraphics[width=#1,keepaspectratio]{#2}\ \\%
        \parbox{3in}{
            \begin{center}
            Fig.\ \arabic{myFCounter}.\ \rm #3
            \end{center}
        }\label{#4}
        \vspace{1ex}
        \end{center}
    \end{minipage}
    \end{center}
    \normalsize
}
\usepackage{bm}

\usepackage{comment}
\excludecomment{long-version}
\includecomment{short-version}

\usepackage{ifpdf}

\newcommand{\esup}{\mathop{\text{ess\;sup}}}    
\renewcommand{\P}{\mathbb{P}}                   
\newcommand{\E}{\mathbb{E}}                     
\newcommand{\Em}{\mathcal{E}}                   
\newcommand{\F}{\mathcal{F}}                    
\newcommand{\Mh}{\mathcal{M}}                      
\newcommand{\R}{\mathbb{R}}                     
\newcommand{\II}[1]{\bm{1}_{\left\{#1\right\}}} 
\newcommand{\bpi}{\pi}                          
\newcommand{\bPi}{\Pi}                          
\newcommand{\e}{e}
\newcommand{\T}{\mathbb{T}}
\newcommand{\M}{\mathbb{M}}                     

\newtheorem{theorem}{Theorem}
\newtheorem{proposition}[theorem]{Proposition}
\newtheorem{lemma}[theorem]{Lemma}

\newtheorem{corollary}[theorem]{Corollary}
\newtheorem{remark}[theorem]{Remark}

\usepackage[unicode]{hyperref}
\hypersetup{
    draft   
}
\usepackage{ifpdf}

\begin{document}

\thispagestyle{empty}

\begin{center}
    \Huge Joint Detection and Identification of an Unobservable Change in the Distribution of a Random Sequence\\
    \vskip 10pt
    \large
    \begin{tabular}[t]{c@{\extracolsep{4em}}c}
        Savas Dayanik and Christian Goulding  & H.\ Vincent Poor \\
        Dept. of Operations Research and Financial Engineering & School of Engineering and Applied Science \\
        Princeton University, Princeton, NJ~~08544 & Princeton University, Princeton, NJ~~08544 \\
        Email: \{sdayanik, cgouldin\}@princeton.edu & Email:  poor@princeton.edu
    \end{tabular}
    \normalsize
\end{center}

\begin{multicols}{2}{
\textbf{\emph{Abstract--}This paper examines the joint problem of
detection and identification of a sudden and unobservable change
in the probability distribution function (pdf) of a sequence of
independent and identically distributed (i.i.d.)\ random variables
to one of finitely many alternative pdf's. The objective is quick
detection of the change and accurate inference of the ensuing pdf.
Following a Bayesian approach, a new sequential decision strategy
for this problem is revealed and is proven optimal.  Geometrical
properties of this strategy are demonstrated via numerical
examples.}

\section{Introduction} \label{sec:Introduction}

Consider a sequence of i.i.d.\ random variables $X_1, X_2,
\ldots$, taking values in some measurable space $(E,\Em)$. The
common probability distribution of the $X$'s is initially some
known probability measure $\P_0$ on $(E,\Em)$, and then, at some
\emph{unobservable} disorder time $\theta$, the common probability
distribution changes suddenly to another probability measure
$\P_{\mu}$ for some \emph{unobservable} index $\mu\in \Mh
\triangleq \{1,\ldots,M\}$. The objective is to detect the change
as quickly as possible, and, at the same time, to identify the new
probability distribution as accurately as possible, so that the
most suitable actions can be taken with the least delay.

This problem can be viewed as the fusion of two fundamental areas
of sequential analysis: change detection and multi-hypothesis
testing.  In traditional change detection problems, there is only
one change distribution, $\P_1$; therefore, the focus is
exclusively on detecting the change time. Whereas, in traditional
sequential multi-hypothesis testing problems, there is no change
time to consider.  Instead, every observation has common
distribution $\P_\mu$ for some unknown $\mu$, and the focus is
exclusively on the inference of $\mu$. Both of these subproblems
have been studied extensively. For recent reviews of these areas,
we refer the reader to \cite{MR1210954} and \cite{MR1844531} and
the references therein.

However, the joint problem involves key trade-off decisions not
taken into account by separately applying techniques for these
subproblems. While raising an alarm as soon as the change occurs
is advantageous for the change detection task, it is undesirable
for the identification task because waiting longer provides more
observations for inferring the change distribution. Likewise, the
unknown change time complicates the identification task, and, as a
result, adaptation of existing sequential multi-hypothesis testing
algorithms is problematic.
\noindent\begin{minipage}[!b]{\columnwidth} \footnotetext{The
research of Savas Dayanik was supported by the Air Force Office of
Scientific Research, under grant AFOSR-FA9550-06-1-0496. The
research of H.\ Vincent Poor was supported in part by the U.S.\
Army Pantheon Project.}
\end{minipage}

Decision strategies for the joint problem have a wide array of
applications, such as fault detection and isolation in industrial
processes, target detection and identification in national
defense, pattern recognition and machine learning, radar and sonar
signal processing, seismology, speech and image processing,
biomedical signal processing, finance, and insurance. However, the
theory has not been broadly developed.
Nikiforov~\cite{Nikiforov1995} provides the first results for this
problem, showing asymptotic optimality for a certain non-Bayesian
approach, and Lai~\cite{Lai2000} generalizes these results through
the development of information-theoretic bounds and the
application of likelihood methods.  In this paper, we follow a
Bayesian approach to reveal a new optimal strategy for this
problem and we describe an accurate numerical scheme for its
implementation.

In Sec.\ \ref{sec:Problem-statement} we formulate precisely the
problem in a Bayesian framework, and in Sec.\
\ref{sec:Reformulation} we show that it can be reduced to an
optimal stopping of a Markov process whose state space is the
standard probability simplex.  In addition, we establish a simple
recursive formula that captures the dynamics of the process and
yields a sufficient statistic fit for online tracking.

In Sec.\ \ref{sec:dynamic-programming-solution} we use optimal
stopping theory to substantiate the optimality equation for the
value function of the optimal stopping problem.  Moreover, we
prove that this value function is bounded, concave, and continuous
on the standard probability simplex and that the optimal stopping
region consists of $M$ non-empty, convex, closed, and bounded
subsets.  Also, we consider a truncated version of the problem
that allows at most $N$ observations from the sequence of random
measurements.  We establish an explicit bound (inversely
proportional to $N$) for the approximation error associated with
this truncated problem.

In Sec.\ \ref{sec:Special-cases} we show that the separate
problems of change detection and sequential multi-hypothesis
testing are solved as special cases of the overall joint solution.
We illustrate some geometrical properties of the optimal method
and demonstrate its implementation by numerical examples for the
special cases $M=2$ and $M=3$.  Specifically, we show instances in
which the $M$ convex subsets comprising the optimal stopping
region are connected and instances in which they are not.
Likewise, we show that the continuation region (i.e., the
complement of the stopping region) need not be connected. We refer
the reader to \cite{DGP06} for complete proofs of the results.

\section{Problem statement}
\label{sec:Problem-statement}

Let $(\Omega,\F, \P)$ be a probability space hosting random
variables $\theta:\Omega\mapsto\{0,1,\ldots\}$ and
$\mu:\Omega\mapsto \Mh \triangleq \{1,\ldots,M\}$ and a process $X
= (X_n)_{n\geq1}$ taking values in some measurable space
$(E,\Em)$. Suppose that for every $t\ge 1$, $i \in \Mh$, $n\ge 1$,
and $(E_k)^n_{k=1}\subseteq \Em $ we have
\begin{multline*}
  \P\left\{\theta=t, \mu=i, X_1 \in E_1,\ldots, X_n \in E_n \right\}  \\
  = (1-p_0) (1-p)^{t-1} p \nu_i \prod_{k=1}^{(t-1) \land n}
  \P_0(E_k) \prod_{\ell = t\lor 1}^{n} \P_i(E_{\ell})
\end{multline*}
for some given probability measures $\P_0,\P_1,\ldots,\P_M$ on
$(E,\Em)$, known constants $p_0\in[0,1]$, $p\in(0,1)$, and
$\nu_i>0,i\in \Mh$ such that $\nu_1+\cdots+\nu_M = 1$, where
$x\wedge y\triangleq \min\{x,y\}$ and $x\vee y\triangleq
\max\{x,y\}$.  Namely, $\theta$ is independent of $\mu$; it has a
zero-modified geometric distribution with parameters $p_0$ and $p$
in the terminology of
\cite[Sec.\
3.6]{MR1490300}, which reduces to the standard geometric
distribution when $p_0=0$.

Conditionally on $\theta$ and $\mu$, the random variables $X_n$, $n\ge
1$ are independent;  $X_1,\ldots, X_{\theta-1}$ and $X_{\theta},
X_{\theta+1},\ldots$ are identically distributed with common
distributions $\P_0$ and $\P_{\mu}$, respectively.  The probability
measures $\P_0,\P_1,\ldots,\P_M$ always admit densities with respect
to some sigma-finite measure $m$ on $(E,\Em)$; for example, we can
take $m = \P_0+\P_1\cdots+\P_M$.  So, we fix $m$ and denote the
corresponding densities by $f_0, f_1,\ldots,f_M$, respectively.

Suppose now that we observe sequentially the random variables
$X_n$, $n\ge 1$. Their common pdf $f_0$ changes at stage $\theta$
to some other pdf $f_{\mu}$, $\mu\in \Mh$. Our objective is to
detect the change time $\theta$ as quickly as possible \emph{and}
to identify the change index $\mu$ as accurately as possible. More
precisely, given costs associated with detection delay, false
alarm, and false identification of the change index, we seek a
strategy that minimizes the expected total change detection
\emph{and} identification cost.

Let $\mathbb{F} = (\F_n)_{n\geq0}$ denote the natural filtration
of the observation process $X$, where
\begin{align*}
  \F_0=\{\varnothing,\Omega\}\quad\text{and}\quad
  \F_n=\sigma(X_1,\ldots,X_n),\quad n\geq1.
\end{align*}
A \emph{strategy} $\delta=(\tau, d)$ is a pair consisting of a
\emph{stopping time} $\tau$ of the filtration $\mathbb{F}$ and a
\emph{terminal decision rule} $d: \Omega \mapsto \Mh$ measurable
with respect to the history $\F_{\tau}=\sigma(X_{n\wedge\tau};
n\geq1)$ of observation process $X$ through stage $\tau$. Applying
a strategy $\delta=(\tau,d)$ consists of announcing at the end of
stage $\tau$ that the common pdf has changed from $f_0$ to $f_d$
at or before stage $\tau$. Let
\begin{align*}
  \Delta \triangleq \{(\tau,d) \mid \tau\in\mathbb{F}, \text{ and
    $d\in\F_\tau$ is an $\Mh$-valued r.\ v.}\}
\end{align*}
denote the collection of all such sequential decision strategies.

For every strategy $\delta=(\tau, d)\in\Delta$, we define a
\emph{Bayes risk function}
\begin{align}
  R(\delta) = c\,\E[(\tau-\theta)^+] + \E[a_{0
    d}\II{\tau<\theta}+a_{\mu
    d}\II{\theta\leq\tau<\infty}]\label{E:BayesRiskUnderP}
\end{align}
\noindent as the expected diagnosis cost: the sum of the expected
detection delay cost and the expected terminal decision cost upon
alarm, where $c>0$ and $a_{ij}\ge 0, i\in\{0\}\cup\Mh,j\in\Mh$ are
known constants satisfying $a_{ii}=0, i\in\Mh$ (i.e., no cost for
a correct terminal decision), and $(x)^+\triangleq\max\{x,0\}$.

The problem is to find a sequential decision strategy
$\delta=(\tau,d)\in\Delta$ (if it exists) with the \emph{minimum
Bayes
  risk}
\begin{align}
  R^* \triangleq \inf_{\delta\in\Delta} R(\delta).\label{E:UDef1}
\end{align}

\section{Posterior analysis and formulation as an optimal stopping
  problem}
\label{sec:Reformulation}

In this section we show that the Bayes risk function in
(\ref{E:BayesRiskUnderP}) can be written as the expected value of the
running and terminal costs driven by a certain Markov process.  We use
this fact to recast the minimum Bayes risk in (\ref{E:UDef1}) as a
Markov optimal stopping problem.

Let us introduce the posterior probability processes
\begin{align*}
  \Pi_n^{(0)} &\triangleq
  \P\{\theta>n\,|\,\F_n\}\quad\text{and}\quad
  \Pi_n^{(i)} \triangleq \P\{\theta\leq n,\mu = i\,|\,\F_n\}
\end{align*}
for $i\in \Mh, n\geq 0$. Having observed the first $n$
observations, $\Pi_n^{(0)}$ is the posterior probability that the
change \emph{has not} yet occurred at or before stage $n$, while
$\Pi_n^{(i)}$ is the posterior joint probability that the change
\emph{has} occurred by stage $n$ and that the hypothesis $\mu=i$
is correct.  The connection of these posterior probabilities to
the loss structure for our problem is established in
the next proposition. 

\begin{proposition}\label{P:BayesRiskInTermsOfPi}
  For every sequential decision strategy $\delta\in\Delta$, the Bayes
  risk function (\ref{E:BayesRiskUnderP}) can be expressed in terms of
  the process $\bPi\triangleq\{ \bPi_n= (\Pi_n^{(0)}, \ldots,
  \Pi_n^{(M)})\}_{n\geq 0}$ as
{
  \begin{align*}
    R(\delta) &= \E\!\left[ \sum_{n=0}^{\tau-1}c\,(1\!-\!\Pi_n^{(0)})
      +\II{\tau<\infty}\!\sum_{j=1}^{M}\II{d=j}\!\sum_{i=0}^{M}
      a_{ij}\Pi_{\tau}^{(i)}\right]\!\!.
  \end{align*}}
\end{proposition}

While our original formulation of the Bayes risk function
(\ref{E:BayesRiskUnderP}) was in terms of the values of the
unobservable random variables $\theta$ and $\mu$, Proposition
\ref{P:BayesRiskInTermsOfPi} gives us an equivalent version of the
Bayes risk function in terms of the posterior distributions for
$\theta$ and $\mu$.  This is particularly effective in light of
Proposition \ref{P:PiProperties}, which we state with the aid of some
additional notation that is referred to throughout the paper.  Let
\begin{align*}
  S^M \triangleq
  \left\{\bpi=(\pi_0,\pi_1,\ldots,\pi_M)\in[0,1]^{M+1}\,\bigm|\,
    {\textstyle\sum_{i=0}^M}\pi_i = 1 \right\}
\end{align*}
denote the standard $M$-dimensional probability simplex. Define
the mappings $D_i:S^M \times E \mapsto [0,1], i\in \Mh$ and $D:S^M
\times E \mapsto [0,1]$ by
\begin{align*}
  D_{i}(\bpi,x) &\triangleq \left\{
    \begin{aligned}
      &(1-p)\pi_0 f_0(x), && i=0\\
      &(\pi_i+\pi_0\,p\nu_i) f_i(x), && i\in \Mh
    \end{aligned}
  \right\}
\end{align*}
and $D(\bpi,x)\triangleq\sum_{i=0}^{M}D_{i}(\bpi,x)$, and the
operator $\T$ on the collection of bounded functions $f:S^M
\mapsto\R$ by
\begin{align}
  \label{E:T-operator}
  (\T f)(\bpi) &\triangleq\!\int_{E}
  m(dx)\,D(\bpi,x)\,f\!\left({\textstyle\frac{D_0(\bpi,x)}{D(\bpi,x)},\ldots,
    \frac{D_M(\bpi,x)}{D(\bpi,x)}}\right)
\end{align}
for every $\bpi\in S^M$.

\begin{proposition}\label{P:PiProperties}
  (a) The process
    $\bPi^{(0)}\triangleq\{\Pi_n^{(0)},\F_n\}_{n\geq 0}$ is a
    supermartingale, and $\E\,\Pi_n^{(0)} \leq (1-p)^n$ for every
    $n\geq 0$.

  (b) The process
    $\bPi^{(i)}\triangleq\{\Pi_n^{(i)},\F_n\}_{n\geq 0}$ is a
    submartingale for every $i\in \Mh$.

  (c) The process
    $\bPi=\{(\Pi_n^{(0)},\ldots,\Pi_n^{(M)})\}_{n\geq 0}$ is a Markov
    process, and
    \begin{align}
      \Pi_{n+1}^{(i)} =
      \frac{D_i(\bPi_n,X_{n+1})}{D(\bPi_n,X_{n+1})},\quad i\in
      \{0\}\cup\Mh ,\quad n\geq 0,\label{E:Pi-Dynamics}
    \end{align}
    with initial state $\Pi_{0}^{(0)} = 1-p_0$ and
    $\Pi_{0}^{(i)}=p_0\nu_i$, $i\in \Mh.$ 
    Moreover, for every bounded function $f:S^M\mapsto\R$ and $n\geq
    0$, we have $\E[f(\bPi_{n+1})|\bPi_n] = (\T f)(\bPi_n)$.
\end{proposition}

\begin{remark}\label{R:PiProperties}
  Since $\bPi$ is uniformly bounded, the limit
  $\lim_{n\rightarrow\infty}\bPi_n$ exists by the martingale
  convergence theorem.  Moreover,
  $\lim_{n\rightarrow\infty}\Pi_n^{(0)}=0$ a.s.\ by Proposition
  \ref{P:PiProperties}(a) since $p\in(0,1)$.
\end{remark}

Now, let the functions $h, h_1,\ldots,h_M$ from $S^M$ into $\R_+$ be
defined by
\begin{align*}
  h(\bpi)\triangleq \min_{j\in \Mh} h_j(\bpi) \quad \text{and} \quad
  h_j(\bpi) \triangleq \sum_{i=0}^{M} \pi_i\, a_{ij},\quad j\in \Mh,
\end{align*}
respectively.  Then, we note that for every $\delta=(\tau,d)\in
\Delta$, we have
\begin{align*}
  R(\tau, d) &= \E\left[ \sum_{n=0}^{\tau-1}c(1-\Pi_n^{(0)})
    +\II{\tau<\infty}\sum_{j=1}^{M}\II{d=j}h_j(\Pi_{\tau})\right]\\
  &\geq \E\left[ \sum_{n=0}^{\tau-1}c(1-\Pi_n^{(0)})
    +\II{\tau<\infty}h(\Pi_{\tau})\right] = R(\tau,\tilde{d})
\end{align*}
where we define on the event $\{\tau<\infty\}$ the terminal decision
rule $\tilde{d}$ to be any index satisfying
$h_{\tilde{d}}(\Pi_{\tau})=h(\Pi_{\tau})$.  In other words, an optimal
terminal decision depends only upon the value of the $\bPi$ process at
the stage in which we stop.  Note also that the functions $h$ and
$h_1,\ldots,h_M$ are bounded on $S^M$.  Therefore, we have the
following:

\begin{lemma}\label{L:OSP1}
  The minimum Bayes risk (\ref{E:UDef1}) reduces to the following
  optimal stopping of the Markov process $\bPi$:
  \begin{align*}
    R^* &= \inf_{(\tau,d)\in\Delta}R(\tau,d) =
    \inf_{(\tau,\tilde{d})\in\Delta}R(\tau,\tilde{d})\\
    &= \inf_{\tau\in\mathbb{F}}
    \,\E\left[\sum_{n=0}^{\tau-1}c\,(1-\Pi_n^{(0)})+\II{\tau<\infty}h(\Pi_\tau)\right].
  \end{align*}
\end{lemma}

We simplify this formulation further by showing that it is enough to
take the infimum over
\begin{align}
  C \triangleq \{\tau\in\mathbb{F}\,|\,\tau<\infty \text{ a.s. and }
  \E Y_\tau^-<\infty\},\label{E:C}
\end{align}
where we define
\begin{align*}
  -Y_n \triangleq \sum_{k=0}^{n-1}c\,(1-\Pi_k^{(0)})+h(\Pi_n),\quad
  n\geq 0
\end{align*}
as the minimum \emph{partial risk} obtained by making the best
terminal decision on $\{\tau=n\}$. Since $h(\cdot)$ is bounded on
$S^M$, the process $\{Y_n, \F_n; n\ge 0\}$ consists of integrable
random variables. So the expectation $\E Y_\tau$ exists for every
$\tau\in\mathbb{F}$, and our problem becomes
\begin{align}
  -R^*=\sup_{\tau\in\mathbb{F}}\E
  Y_\tau.\label{E:OptimizationProblemTau}
\end{align}

Observe that $\E \tau <\infty$ for every $\tau\in C$ because $\infty >
(1/c)\E Y_\tau^- \geq \E (\tau-\theta)^+ \geq \E (\tau-\theta) \geq \E
\tau -\E \theta \ge \E\tau - (1/p)$.  In fact, we have $\E
Y_\tau>-\infty \Leftrightarrow \E Y_\tau^-<\infty \Leftrightarrow \E
\tau<\infty$ for every $\tau\in\mathbb{F}$.  Since
$\sup_{\tau\in\mathbb{F}}\E Y_{\tau} \geq \E Y_0 > -h(\Pi_0) >
-\infty$, it is enough to consider $\tau\in\mathbb{F}$ such that
$\E\tau <\infty$. Namely, (\ref{E:OptimizationProblemTau}) reduces to
\begin{align}
  -R^*=\sup_{\tau\in C} \E Y_\tau.\label{E:OptimizationProblemC}
\end{align}

\section{Solution via optimal stopping theory}
\label{sec:dynamic-programming-solution}

In this section we derive an optimal solution for the problem
in~\eqref{E:UDef1} by building on the formulation
of~\eqref{E:OptimizationProblemC} via the tools of optimal
stopping theory, which are detailed in \cite{MR0331675}.

\mysubsection{The optimality equation}\label{sec:Derive-Opt-Eqn}

We begin by applying the method of truncation with a view of
passing to the limit to arrive at the final result.  Define for
every pair of integers $n, N$ satisfying $0 \le n \le N$ the
sub-collections
\begin{align*}
  C_n &\triangleq \{\tau \vee n\,|\,\tau\in C\}\quad\text{and}\quad
  C_n^N \triangleq \{\tau \wedge N\,|\,\tau\in C_n\}
\end{align*}
of stopping times in $C$ of \eqref{E:C} and the families of
(truncated) optimal stopping problems
\begin{align}
  \label{E:Vn-and-VnN}
  -V_n \triangleq \sup_{\tau\in C_n}\E Y_\tau
  \quad\text{and}\quad
  -V_n^N \triangleq \sup_{\tau\in C_n^N}\E
  Y_\tau
\end{align}
corresponding to $(C_n)_{n\geq 0}$ and $(C_n^N)_{0\leq n\leq N}$,
respectively. Note that $C\equiv C_0$ and $R^*\equiv V_0$.

To investigate these optimal stopping problems, we introduce
versions of the \emph{Snell envelope} of $(Y_n)_{n\geq 0}$ (i.e.,
the smallest regular supermartingale dominating $(Y_n)_{n\geq
  0}$) corresponding to $(C_n)_{n\geq 0}$ and $(C_n^N)_{0\leq n\leq
  N}$, respectively, defined by
\begin{align*}
  \gamma_n &\triangleq \esup_{\tau\in C_n} \E [Y_\tau\,|\,\F_n]
  \quad\text{and}\quad \gamma_n^N \triangleq \esup_{\substack{\tau\in
      C_n^N}} \E [Y_\tau\,|\,\F_n].
\end{align*}
Then through the following series of lemmas we point out several
useful properties of these Snell envelopes. Finally, we extend
these results to an arbitrary initial state vector and establish
the optimality equation.  Note that each of the ensuing
(in)equalities between random variables are in the $\P$-almost
sure sense.

First, these Snell envelopes provide the following alternative
expressions for the optimal stopping problems introduced in
\eqref{E:Vn-and-VnN} above.

\begin{lemma}\label{L:Vn-equal-expected-gamma}
  For every $N\ge 0$ and $0\le n\le N$, we have $-V_n = \E \gamma_n$
  and $-V_n^N = \E \gamma_n^N$.
\end{lemma}

Second, we have the following backward-induction equations.

\begin{lemma}\label{L:backward-induction-eqns}
  We have $\gamma_n = \max\{Y_n, \E [\gamma_{n+1}\,|\,\F_n]\}$ for
  every $n\ge 0$. For every $N\ge 1$ and $0\le n \le N-1$, we have
  $\gamma_N^N = Y_N$ and $\gamma_n^N = \max\{Y_n, \E
  [\gamma_{n+1}^N\,|\,\F_n]\}$.
\end{lemma}

We also have that these versions of the Snell envelopes coincide in
the limit as $N\rightarrow\infty$.  That is,

\begin{lemma}\label{L:gamma-equals-gamma-prime}
  For every $n\geq 0$, we have $\gamma_n = \lim_{N\rightarrow\infty}
  \gamma_n^N$.
\end{lemma}

Next, recall from \eqref{E:T-operator} and Proposition
\ref{P:PiProperties}(c) the operator $\T$ and let us introduce the
operator $\M$ on the collection of bounded functions $f:S^M
\mapsto \R_+$ defined by
\begin{align*}
  (\M f)(\bpi) \triangleq
  \min\{h(\bpi),c(1-\pi_0)+(\T f)(\bpi)\},\quad\bpi\in S^M.
\end{align*}
Observe that $0\leq \M f \leq h$.  That is, $\bpi\mapsto(\M f)(\bpi)$
is a nonnegative bounded function. Therefore, $\M^2 f\equiv \M(\M f)$
is well-defined.  If $f$ is nonnegative and bounded, then $\M^n
f\equiv \M(\M^{n-1} f)$ is defined for every $n\ge 1$, with
$\M^0 f\equiv f$ by definition.  Using operator $\M$, we can express
$(\gamma_n^N)_{0\leq n\leq N}$ in terms of the process $\bPi$ as
stated in the following lemma.

\begin{lemma}\label{L:pg-36}
  For every $N\ge 0$, and $0\le n \le N$, we have
    $\gamma_n^N = -c\sum_{k=0}^{n-1}(1-\Pi_k^{(0)})-(\M^{N-n}h)(\Pi_n)$.
\end{lemma}

The next lemma shows how the optimal stopping problems can be
rewritten in terms of the operator $\M$. It also conveys the
connection between the truncated optimal stopping problems and the
initial state $\bPi_0$ of the $\bPi$ process.

\begin{lemma}\label{L:pg-38}
  We have
  (a) $V_0^N=(\M^N h)(\bPi_0)$ for every $N\geq 0$, and
  (b) $V_0={\displaystyle\lim_{N\rightarrow\infty}(\M^N
      h)(\bPi_0)}$.
\end{lemma}

Observe that since $\bPi_0\in\F_0=\{\varnothing,\Omega\}$, we have
$\P\{\bPi_0=\bpi\}=1$ for some $\bpi\in S^M$.  On the other hand, for
every $\bpi\in S^M$ we can construct a probability space
$(\Omega,\F,\P_{\bpi})$ hosting a Markov process $\bPi$ with the same
dynamics as in \eqref{E:Pi-Dynamics} and $\P_{\bpi}\{\bPi_0=\bpi\}=1$.
Moreover, on such a probability space, the preceding results remain
valid.  So, let us denote by $\E_{\bpi}$ the expectation with respect
to $\P_{\bpi}$ and rewrite \eqref{E:Vn-and-VnN} as
\begin{align*}
  -V_n(\bpi) \triangleq \sup_{\tau\in C_n}\E_{\bpi} Y_\tau
  \quad \text{and} \quad -V_n^N(\bpi) \triangleq \sup_{\tau\in
    C_n^N}\E_{\bpi} Y_\tau
\end{align*}
for every $\bpi\in S^M$.  Then Lemma \ref{L:pg-38} implies that
\begin{align}
\label{eq:value-functions}
  V_0^N\!(\bpi)=(\M^N h)(\bpi)\!\!\quad\text{ and
  }\quad\!\!V_0(\bpi)=\lim_{N\rightarrow\infty}(\M^N h)(\bpi)
\end{align}
for every $\bpi\in S^M$.  Taking limits as $N\rightarrow\infty$ of
both sides in $(\M^{N+1}h)(\bpi) = \M(\M^N h)(\bpi)$ and applying
the monotone convergence theorem on the right-hand side yields
$V_0(\bpi) = (\M V_0)(\bpi)$.  Hence, we have shown the following
result.

\begin{proposition}[Optimality equation]\label{P:Dyn-prog-eqn}
  For every $\bpi\in S^M$, 
  \begin{align}
    V_0(\bpi)\!=\!(\M V_0)(\bpi) \equiv
    \min\{h(\bpi),c(1\!-\!\pi_0)\!+\!(\T V_0)(\bpi)\}.\label{E:Dyn-prog-eqn}
  \end{align}
\end{proposition}

\begin{remark}
  By solving $V_0(\bpi)$ for any initial state $\bpi\in S^M$, we
  capture the solution to the original problem since property (c) of
  Proposition \ref{P:PiProperties} and \eqref{E:OptimizationProblemC}
  imply that
    $R^* = V_0(1-p_0,p_0\nu_1,\ldots,p_0\nu_M)$.
\end{remark}

\mysubsection{Some properties of the value
function}\label{sec:V-properties}

Now, we reveal some important properties of the value function
$V_0(\cdot)$ of (\ref{eq:value-functions}).  These results help us
to establish an optimal solution for $V_0(\cdot)$, and hence an
optimal solution for $R^*$, in the next subsection.

\begin{lemma}\label{L:V-concave}
    If $g:S^M \mapsto \R$ is a bounded concave function, then so is $\T g$.
\end{lemma}

\begin{proposition}\label{P:V-concave}
  The mappings $\bpi \mapsto V_0^N(\bpi), N\geq 0$ and $\bpi \mapsto
  V_0(\bpi)$ are concave.
\end{proposition}

\begin{proposition}\label{P:V-convergence-rate}
    For every $N\ge 1$ and $\bpi\in S^M$, we have
        \begin{align*}
            V_0(\bpi)\leq V_0^N(\bpi) \leq
            V_0(\bpi)+\left(\frac{\|h\|^2}{c}+\frac{\|h\|}{p}\right)\frac{1}{N}.
        \end{align*}
        Since $\|h\|\triangleq \sup_{\bpi\in S^M} |h(\bpi)|<\infty$,
        $\lim_{N\rightarrow\infty} \downarrow V_0^N(\bpi) = V_0(\bpi)$
        uniformly in $\bpi\in S^M$.
\end{proposition}

\begin{proposition}\label{P:V0N-continuous}
  For every $N\ge 0$, the function $V_0^N:S^M\mapsto\R_+$ is
  continuous.
\end{proposition}

\begin{corollary}\label{C:V-continuous}
  The function $V_0:S^M \mapsto \R_+$ is continuous.
\end{corollary}

Note that $S^M$ is a compact subset of $\R^{M+1}$, so while continuity
of $V_0(\cdot)$ on the interior of $S^M$ follows from the concavity of
$V_0(\cdot)$ by Proposition \ref{L:V-concave}, Corollary
\ref{C:V-continuous} establishes continuity on all of $S^M$, including
its boundary.

\mysubsection{An optimal sequential decision
  strategy}\label{sec:optimal-soln}

Finally, we describe the optimal stopping region in $S^M$ implied
by the value function $V_0(\cdot)$, and we present an optimal
sequential decision strategy for our problem. Let us define for
every $N\ge 0$,
\begin{align*}
  \Gamma_N &\triangleq \{\bpi\in S^M\,|\, V_0^N(\bpi)=h(\bpi)\},\\
  \Gamma_N^{(j)} &\triangleq \Gamma_N \cap \{\bpi\in S^M\,|\,
  h(\bpi)=h_j(\bpi)\}, \; j\in \Mh, \\
  \Gamma &\triangleq \{\bpi\in S^M\,|\, V_0(\bpi)=h(\bpi)\},\\
  \Gamma^{(j)} &\triangleq \Gamma \cap \{\bpi\in S^M\,|\,
  h(\bpi)=h_j(\bpi)\}, \; j\in \Mh.
\end{align*}
For each $j\in \{0\}\cup \Mh$, let $\e_j\in S^M$ denote the unit
vector consisting of zero in every component except for the $j$th
component, which is equal to one. Note that
$\e_0,\ldots,\e_M$ are the extreme points of the closed
convex set $S^M$, and any vector $\bpi=(\pi_0,\ldots,\pi_M)\in S^M$
can be expressed in terms of $\e_0,\ldots,\e_M$ as $\bpi =
\sum_{j=0}^{M}\pi_j\e_j$.

\begin{theorem}\label{T:Gamma-decreasing-subsets}
  For every $j\in \Mh$, $(\Gamma_N^{(j)})_{N\geq 0}$ is a decreasing
  sequence of non-empty, closed, convex subsets of $S^M$.  Moreover,
  \begin{gather*}
    \Gamma_0^{(j)} \supseteq \Gamma_1^{(j)} \supseteq \cdots \supseteq
    \Gamma^{(j)},\\
    \Gamma^{(j)}\supseteq \left\{\bpi\in S^M
      \,|\,h_j(\bpi)\leq\min\{h(\bpi),c(1-\pi_0)\}\right\} \ni
    \e_j,\\
    \Gamma = \bigcap_{N=1}^{\infty}\Gamma_N =
    \bigcup_{j=1}^{M}\Gamma^{(j)},\quad\text{and}\quad
    \Gamma^{(j)}=\bigcap_{N=1}^{\infty}\Gamma_N^{(j)},\quad
    j\in \Mh.
  \end{gather*}
  Furthermore, $S^M = \Gamma_0 \supseteq \Gamma_1 \supseteq \cdots
  \supseteq \Gamma \supsetneqq \{\e_1,\ldots,\e_M\}$.
\end{theorem}

\begin{lemma}\label{L:gamma-n-V}
  For every $n\geq 0$, we have $\gamma_n =
  -c\sum_{k=0}^{n-1}(1-\Pi_k^{(0)})-V_0(\Pi_n).$
\end{lemma}

\begin{theorem}\label{T:sigma-properties}
  Let $\sigma \triangleq \inf\{n\geq 0 \,|\, \bPi_n \in\Gamma\}$.
  (a) The stopped process $\{\gamma_{n \wedge\sigma}, \F_n;
    n\geq 0\}$ is a martingale.

  (b) The random variable $\sigma$ is an optimal stopping time
    for $V_0$, and

  (c) $\E\,\sigma<\infty$.
\end{theorem}

Therefore, the pair $(\sigma, d^*)$ is an optimal sequential
decision strategy for \eqref{E:UDef1}, where the optimal stopping
rule $\sigma$ is given by Theorem~\ref{T:sigma-properties}, and,
as in the proof of Lemma~\ref{L:OSP1}, the optimal terminal
decision rule $d^*$ is given by
\begin{align*}
  d^* = j\! \quad\! \text{ on the event}\! \quad\! \{\sigma=n, \bPi_n\in
  \Gamma^{(j)}\}\! \quad\! \text{ for every } n\geq 0.
\end{align*}
Accordingly, the set $\Gamma$ is called the \emph{stopping region}
implied by $V_0(\cdot)$, and
Theorem~\ref{T:Gamma-decreasing-subsets} reveals its basic
structure.  We demonstrate the use of these results in the
numerical examples of Sec.\ \ref{sec:Special-cases}.

Note that we can take a similar approach to prove that the
stopping rules $\sigma_N\triangleq\inf\{n\geq 0\,|\, \bPi_n \in
\Gamma_{N-n}\}, N\geq 0$ are optimal for the truncated problems
$V_0^N(\cdot), N\geq 0$ in (\ref{eq:value-functions}).  Thus, for
each $N\geq 0$, the set $\Gamma_{N}$ is called the stopping region
for $V_0^N(\cdot)$: it is optimal to terminate the experiments in
$\Gamma_N$ if $N$ stages are left before truncation.

\section{Special cases and examples}\label{sec:Special-cases}

\mysubsection{A.\ N.\ Shiryaev's sequential change detection
problem}

Set $a_{0j}=1$ for $j\in \Mh$ and $a_{ij}=0$ for $i,j\in \Mh$, then
the Bayes risk function \eqref{E:BayesRiskUnderP} becomes
  $R(\delta) = \P\{\tau<\theta\} + c\,\E[(\tau-\theta)^+]$.
This is the Bayes risk studied by Shiryaev
\cite{MR0155708,MR0468067} to solve the sequential change
detection problem.

\mysubsection{Sequential multi-hypothesis testing}

Set $p_0=1$, then $\theta = 0$ a.s.\ and thus the Bayes risk
function \eqref{E:BayesRiskUnderP} becomes
  $R(\delta) = \E[c\tau + a_{\mu d}\II{\tau<\infty}]$.
This gives the sequential multi-hypothesis testing problem studied
by Wald and Wolfowitz \cite{MR0034005} and Arrow, Blackwell, and
Girshick \cite{MR0032173}; see also \cite{MR597146}.

\mysubsection{Two alternatives after the change}

In this subsection we consider the special case $M=2$ in which we
have only two possible change distributions, $f_1(\cdot)$ and
$f_2(\cdot)$. We describe a graphical representation of the
stopping and continuation regions for an arbitrary instance of the
special case $M=2$.  Then we use this representation to illustrate
geometrical properties of the optimal method (Sec.\
\ref{sec:dynamic-programming-solution}.\ref{sec:optimal-soln}) via
model instances for certain choices of the model parameters $p_0$,
$p$, $\nu_1$, $\nu_2$, $f_0(\cdot)$, $f_1(\cdot)$, $f_2(\cdot)$,
$a_{01}$, $a_{02}$, $a_{12}$, $a_{21}$, and $c$.

Let the linear mapping $L:\R^3\mapsto\R^2$ be defined by
$L(\pi_0,\pi_1,\pi_2)\triangleq(\tfrac{2}{\sqrt{3}}\pi_1
+\tfrac{1}{\sqrt{3}}\pi_2,\pi_2)$. Since $\pi_0=1-\pi_1-\pi_2$ for
every $\bpi=(\pi_0,\pi_1,\pi_2)\in S^2\subset\R^3$, we can recover
the preimage $\bpi$ of any point $L(\bpi)\in L(S^2)\subset\R^2$.
For every point $\bpi=(\pi_0,\pi_1,\pi_2)\in S^2$, the coordinate
$\pi_i$ is given by the Euclidean distance from the image point
$L(\bpi)$ to the edge of the image triangle $L(S^2)$ that is
\emph{opposite} the image point $L(\e_i)$, for each $i=0,1,2$. For
example, the distance from the image point $L(\bpi)$ to the edge
of the image triangle opposite the lower-left-hand corner
$L(1,0,0)=(0,0)$ is the value of the preimage coordinate $\pi_0$.
See Fig.\ \ref{F:S2-to-2D}.

\ifpdf
\myFigure{3.5in}{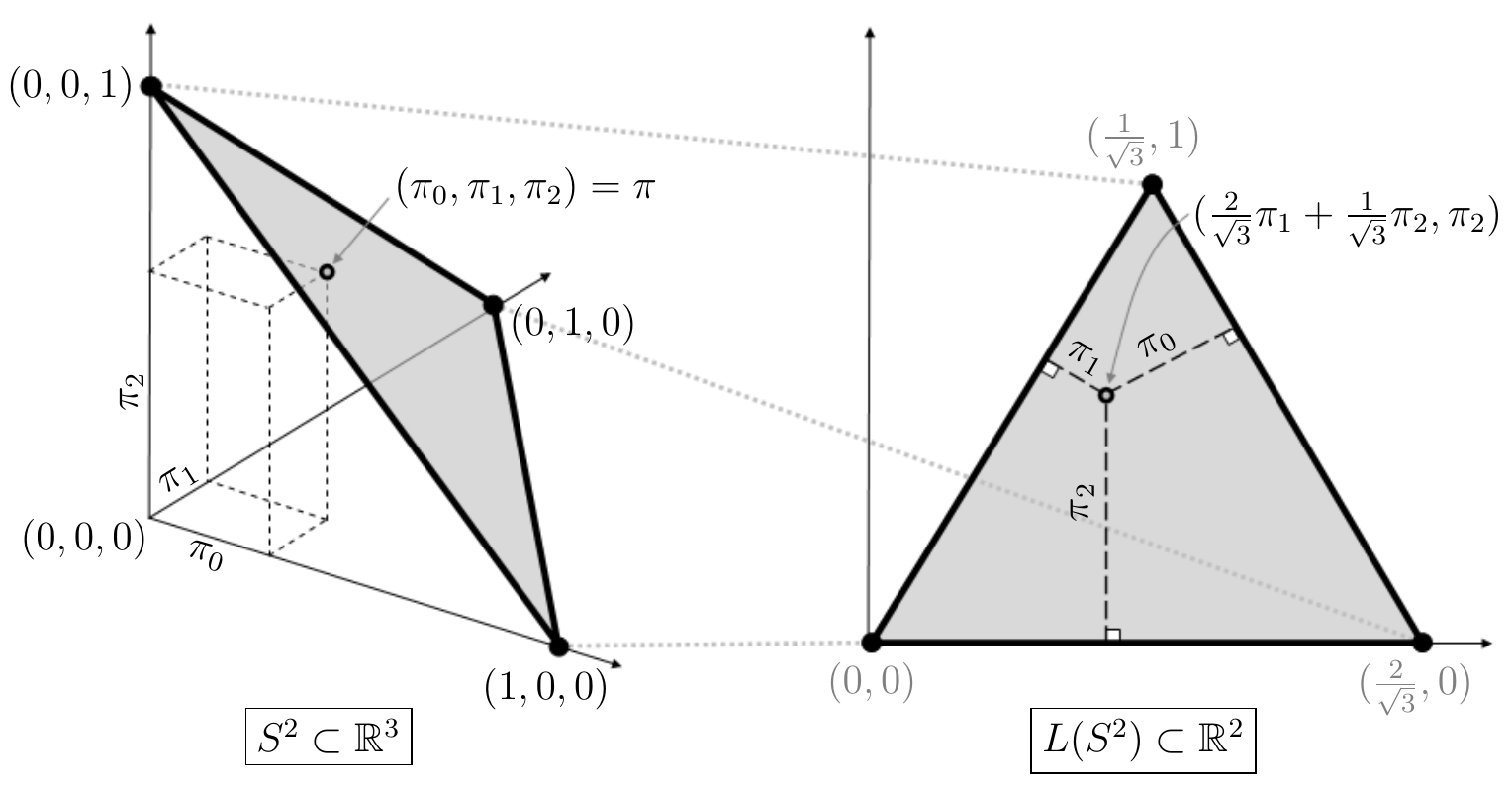}{Linear mapping $L$ of the
standard two-dimensional probability simplex $S^2$ from the
positive orthant of $\R^3$ into the positive quadrant of
$\R^2$.}{F:S2-to-2D}
\fi

Therefore, we can work with the mappings $L(\Gamma)$ and
$L(S^2\setminus\Gamma)$ of the stopping region $\Gamma$ and the
continuation region $S^2\setminus\Gamma$, respectively.
Accordingly, we depict the decision region for each instance in
this subsection using the two-dimensional representation as in the
right-hand-side of Fig.\ \ref{F:S2-to-2D} and we drop the
$L(\cdot)$ notation when labeling various parts of each figure to
emphasize their source in $S^2$.

Each of the examples in this section have the following model
parameters in common:
\begin{gather*}
  p_0=\tfrac{1}{50},\quad p=\tfrac{1}{20},\quad
  \nu_1=\nu_2=\tfrac{1}{2},\\
  f_0\!=\!\left(\tfrac{1}{4}, \tfrac{1}{4}, \tfrac{1}{4},
    \tfrac{1}{4}\right)\!,
  f_1\!=\!\left(\tfrac{4}{10}, \tfrac{3}{10}, \tfrac{2}{10},
    \tfrac{1}{10}\right)\!,
  f_2\!=\!\left(\tfrac{1}{10}, \tfrac{2}{10}, \tfrac{3}{10},
    \tfrac{4}{10}\right)\!.
\end{gather*}
We vary the delay cost and false alarm/identification costs to
illustrate certain geometrical properties of the continuation and
stopping regions.  See Figs.\ \ref{F:2D1}, \ref{F:2D2}, and
\ref{F:2D3}.

\ifpdf
\myFigure{3.5in}{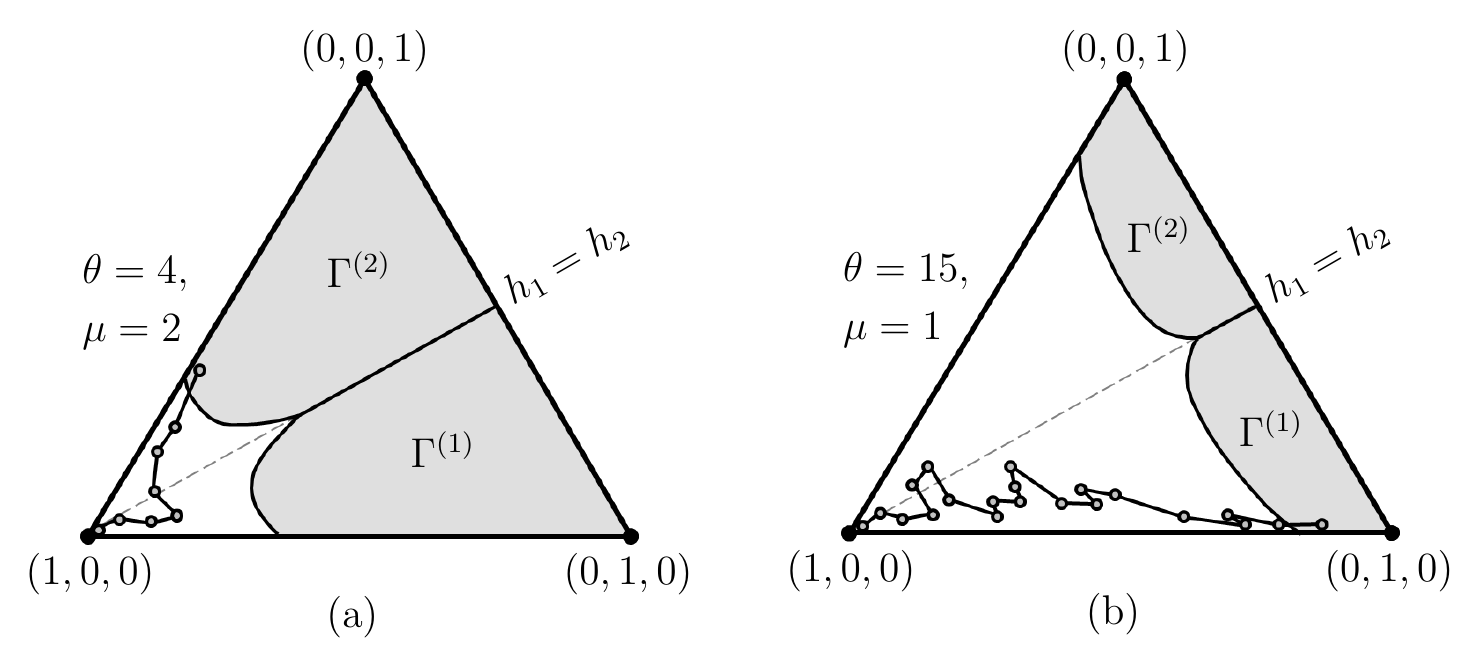}{Illustration of
\emph{connected} stopping regions and the effects of false-alarm
costs. (a) and (b): $a_{12}=a_{21}=3,\,c=1$. (a):
$a_{01}=a_{02}=10$. (b): $a_{01}=a_{02}=50$.}{F:2D1}
\fi

\ifpdf
\myFigure{3.5in}{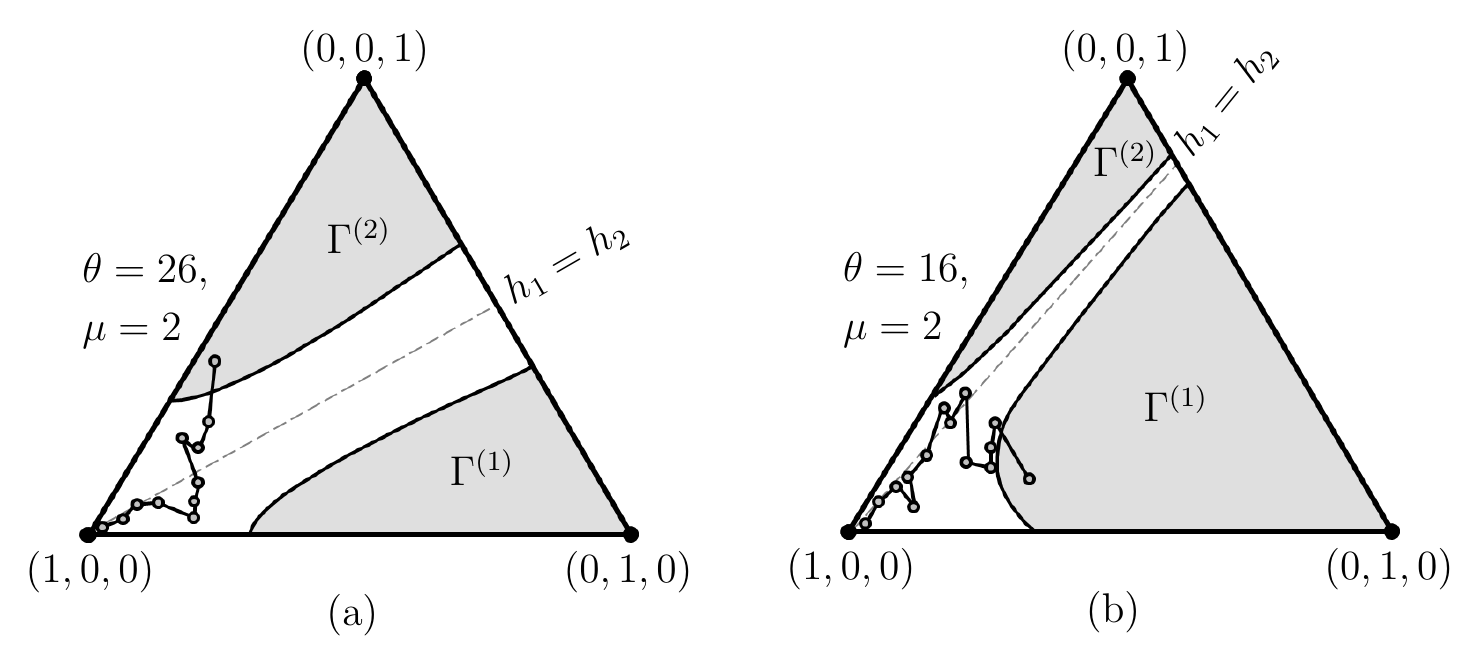}{Illustration of
\emph{disconnected} stopping regions and the effects of asymmetric
false-identification costs. (a) and (b): $a_{01}=a_{02}=10,\,c=1$.
(a): $a_{12}=a_{21}=10$. (b): $a_{12}=16,a_{21}=4$.}{F:2D2}
\fi

\ifpdf
\myFigure{3.5in}{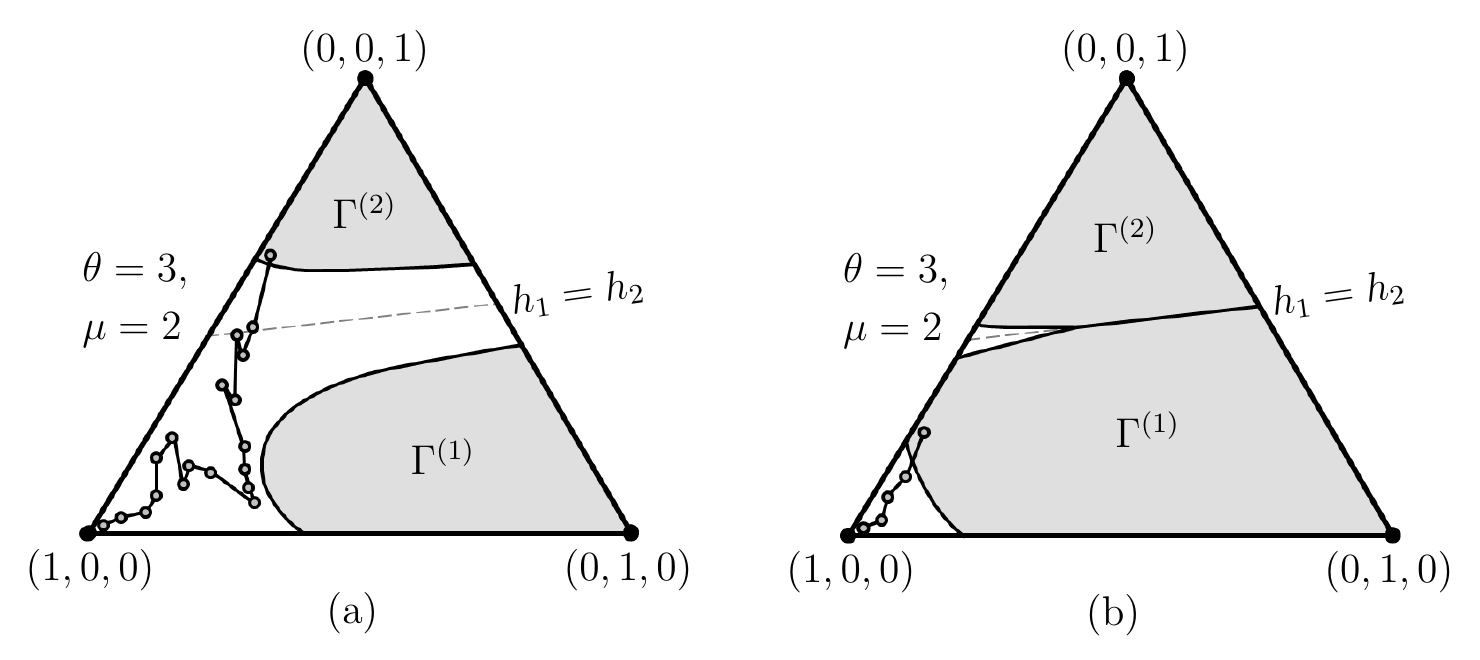}{Illustration of a
\emph{disconnected} continuation region  and the effects of
variation in the delay cost. (a) and (b):
$a_{01}=14,a_{02}=20,a_{12}=a_{21}=8$. (a): $c=1$.  (b):
$c=2$.}{F:2D3}
\fi

These figures have certain features in common. On each subfigure
there is a dashed line representing those states $\bpi\in S^2$ at
which $h_1(\bpi)=h_2(\bpi)$.  Also, each subfigure shows a sample
path of $(\bPi_n)_{n=0}^{\sigma}$ and the realizations of $\theta$
and $\mu$ for the sample.  The shaded area, including its solid
boundary, represents the optimal stopping region, while the
unshaded area represents the continuation region.

Specifically, these figures show instances in which the $M=2$
convex subsets comprising the optimal stopping region are
connected (Fig.\ \ref{F:2D1}) and instances in which they are not
(Figs.\ \ref{F:2D2} and \ref{F:2D3}(a)). Fig.\ \ref{F:2D3}(b)
shows an instance in which the continuation region is
disconnected.

An implementation of the optimal strategy as described in Sec.\
\ref{sec:dynamic-programming-solution}.\ref{sec:optimal-soln} is
as follows: Initialize the statistic $\bPi=(\bPi_n)_{n\geq 0}$ by
setting $\bPi_0=(1-p_0,p_0\nu_1,p_0\nu_2)$ as in part (c) of
Proposition \ref{P:PiProperties}. Use the dynamics of
\eqref{E:Pi-Dynamics} to update the statistic $\bPi_n$ as each
observation $X_n$ is realized. Stop taking observations when the
statistic $\bPi_n$ enters the stopping region
$\Gamma=\Gamma^{(1)}\cup\Gamma^{(2)}$ for the first time, possibly
before the first observation is taken (i.e., $n=0$). The optimal
terminal decision is based upon whether the statistic $\bPi_n$ is
in $\Gamma^{(1)}$ or $\Gamma^{(2)}$ upon stopping. Each of the
sample paths in Figs.\ \ref{F:2D1}, \ref{F:2D2}, and \ref{F:2D3}
were generated via this algorithm. As Fig.\ \ref{F:2D1} shows, the
sets $\Gamma^{(1)}$ and $\Gamma^{(2)}$ can intersect on their
boundaries and so it is possible to stop in their intersection. In
this case, either of the decisions $d=1$ or $d=2$ is optimal.

We use value iteration of the optimality
equation~\eqref{E:Dyn-prog-eqn} over a fine discretization of
$S^2$ to compute $V_0(\cdot)$ and generate the decision region for
each subfigure.  The resulting discretized decision region is
mapped into the plane via $L$.  See~\cite[Ch.\ 3]{MR2182753} for
techniques of computing the value function via the optimality
equation such as value iteration.

\mysubsection{Three alternatives after the change}

In this subsection we consider the special case $M=3$ in which we
have three possible change distributions, $f_1(\cdot)$,
$f_2(\cdot)$, and $f_3(\cdot)$. Here, the continuation and
stopping regions are subsets of $S^3\subset\R^4$. Similar to the
two-alternatives case, we introduce the mapping of
$S^3\subset\R^4$ into $\R^3$ via
$(\pi_0,\pi_1,\pi_2,\pi_3)\mapsto$
    \begin{align*}
        {\textstyle
        \left(\sqrt{\tfrac{3}{2}}\pi_1
        +\tfrac{1}{2}\sqrt{\tfrac{3}{2}}\pi_2
        +\tfrac{1}{2}\sqrt{\tfrac{3}{2}}\pi_3,
        \tfrac{3}{2}\sqrt{\tfrac{1}{2}}\pi_2 +
        \tfrac{1}{2}\sqrt{\tfrac{1}{2}}\pi_3, \pi_3\right)}.
    \end{align*}
Then we use this representation---actually a rotation of it---to
illustrate in Fig.\ \ref{F:3D} an instance with the following
model parameters:
    \begin{gather*}
        p_0=\tfrac{1}{50},\quad p=\tfrac{1}{20},\quad
        \nu_1=\nu_2=\nu_3=\tfrac{1}{3}, \\
        f_0=\left(\tfrac{1}{4}, \tfrac{1}{4}, \tfrac{1}{4},
        \tfrac{1}{4}\right),\quad
        f_1=\left(\tfrac{4}{10}, \tfrac{3}{10}, \tfrac{2}{10},
        \tfrac{1}{10}\right),\\
        f_2=\left(\tfrac{1}{10}, \tfrac{2}{10}, \tfrac{3}{10},
        \tfrac{4}{10}\right),\quad
        f_3=\left(\tfrac{3}{10}, \tfrac{2}{10}, \tfrac{2}{10},
        \tfrac{3}{10}\right),\\
        c=1,\quad a_{0j}=40,\quad a_{ij}=20,\quad i,j=1,2,3.
    \end{gather*}

Fig.\ \ref{F:3D} can be interpreted in a manner similar to the
figures of the previous subsection.  In this case, for every point
$\bpi=(\pi_0,\pi_1,\pi_2,\pi_3)\in S^3$, the coordinate $\pi_i$ is
given by the (Euclidean) distance from the image point $L(\bpi)$
to the face of
\begin{minipage}{3.5in} \ifpdf \myFigure{3.5in}{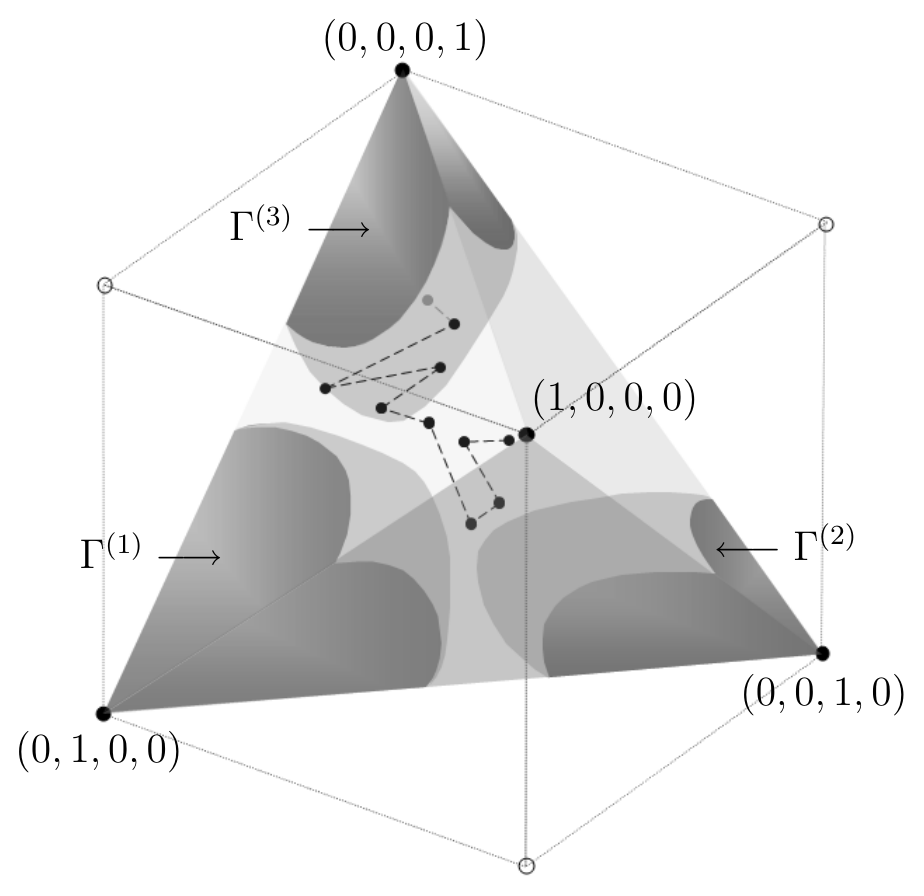}{Illustration of the optimal
decision region for an instance of the special case $M=3$. A
sample path of the process $\bPi$ is shown for which $\theta=6$
and $\mu=3$.}{F:3D} \fi\text{ }\\
\end{minipage}
the image tetrahedron $L(S^3)$ that is opposite the image corner
$L(\e_i)$, for each $i=0,1,2,3$.

\vskip 15pt

\bibliography{82a}
\bibliographystyle{unsrt}
}
\end{multicols}

\end{document}